\newcount\style
\ifodd\style
  \documentstyle[aas2pp4]{article}
  \input psfig
\else
  \documentstyle[12pt,aasms4]{article}
\fi

\begin{document}

\def\om{\Omega_p}
\def\len{a_B}
\def\lag{D_L}
\def\ml{M/L}
\def\kms{\hbox{$\rm {km}~\rm s^{-1}$}}
\def\spose#1{\hbox to 0pt{#1\hss}}
\def\ltsim{\mathrel{\spose{\lower.5ex\hbox{$\mathchar"218$}}
     \raise.4ex\hbox{$\mathchar"13C$}}}
\def\gtsim{\mathrel{\spose{\lower.5ex \hbox{$\mathchar"218$}}
     \raise.4ex\hbox{$\mathchar"13E$}}}
\def\ptrsla{Phil. Trans. Roy. Soc. Lon. A}

\title{Dynamical Friction and the Distribution of Dark Matter in Barred 
Galaxies}

\author{Victor P.\ Debattista and J.\ A.\ Sellwood}
\affil{Department of Physics and Astronomy, Rutgers University, Box 0849 
Piscataway, NJ 08855-0849}
\affil{debattis@physics.rutgers.edu}

\begin{abstract}
We use fully self-consistent $N$-body simulations of barred galaxies to show 
that dynamical friction from a dense dark matter halo dramatically slows the 
rotation rate of bars.  Our result supports previous theoretical predictions for 
a bar rotating within a massive halo.  On the other hand, low density halos, 
such as those required for maximum disks, allow the bar to continue to rotate at 
a high rate.  There is somewhat meager observational evidence indicating that 
bars in real galaxies do rotate rapidly and we use our result to argue that dark 
matter halos must have a low central density in all high surface brightness disk 
galaxies, including the Milky Way.  Bars in galaxies that have larger fractions 
of dark matter should rotate slowly, and we suggest that a promising place to 
look for such candidate objects is among galaxies of intermediate surface 
brightness.
\end{abstract}

\section{Introduction}
The failure of disk galaxy rotation curves to decline significantly at large 
radii (e.g., Rubin et al.\ 1980; Bosma 1981) is usually interpreted as evidence 
that the luminous disk is embedded within a more extensive halo of dark matter 
(DM).  Decomposition of rotation curves into separate contributions from 
luminous and dark matter is an issue of some controversy, however.

One generally assigns a constant mass-to-light ratio, $\ml$, to the disk (and 
bulge, if treated separately) and varies parameters of the assumed DM 
distribution to obtain the best fit to the observed rotation curve.  The 
controversy has arisen because this procedure is very poorly constrained.  The 
$\ml$ values of the luminous matter are uncertain and the fitted functional form 
for the halo is generally chosen arbitrarily (e.g., Broeils 1992, but see also 
Navarro et al.\ 1996).  Moreover, few galaxies manifest a feature in the shape 
of the rotation curve to mark the radius at which the dominant contribution to 
the central attraction switches from luminous to dark matter.  Some argue 
that the DM contribution is substantial at all radii (e.g., van der Kruit 1995), 
while others (e.g., 
van Albada \& Sancisi 1986, Freeman 1992) argue for the so-called maximum disk 
hypothesis, which requires the largest possible mass in luminous matter 
consistent with not exceeding the inner rotation curve.  Maximum disks are 
motivated by the fact that luminous matter alone generally accounts well for 
overall shapes of rotation curves in the inner parts of bright galaxies (Kalnajs 
1983, Kent 1986, Buchhorn 1992, Palunas 1996).

In this {\it Letter}, we introduce a new argument, quite separate from rotation 
curve fitting, to bear on the question of the DM content of disk galaxies: a 
rapidly rotating bar within a massive halo must lose angular momentum to the 
halo through dynamical friction.  Sellwood (1980) reported a secular gain of 
angular momentum by the halo in a low-quality simulation of a barred galaxy, and 
this effect had been predicted by Tremaine (c 1975, unpublished).  Weinberg 
(1985) estimated the bar deceleration rate using linear perturbative 
calculations and simple models for the bar and halo, finding that the bar would 
slow down on the surprisingly short time scale of a few rotation periods.  Until 
recently, this prediction had not been thoroughly tested: simulations by Combes 
et al.\ (1990) had only a bulge of live particles, not an extensive halo; Raha 
et al.\ (1991) did not evolve their models for long enough; the ``halo'' 
particles employed by Little \& Carlberg (1991) were confined to a plane, and 
the bar used by Hernquist \& Weinberg (1992) was rigid and their model had no 
disc.  Our realistic simulations (Debattista \& Sellwood 1996), and those of 
Athanassoula (1996), have confirmed that self-consistent bars are rapidly braked 
by a massive responsive halo.

In order to relate theoretical work to real galaxies, we use the dimensionless 
ratio $\lag/\len$ to characterize the rate of rotation of a bar.  Here $\lag$ is 
the distance from the center of the galaxy to the Lagrange point on the bar 
major axis (corotation) and $\len$ is the semi-major axis of the bar.  This 
ratio is expected to exceed unity in self-consistent stellar bars (Contopoulos 
1980) and is found to be not much larger for the bars that form in most $N$-body 
simulations of bar unstable disks (e.g., Sellwood 1981; Combes \& Sanders 1981). 
 We describe as slow bars those for which $\lag/\len$ is substantially greater 
than unity.

The determination of this parameter for galaxies ideally requires a direct 
measurement of the bar pattern speed, $\om$.  Merrifield \& Kuijken (1995) 
estimate $\lag/\len = 1.4 \pm 0.3$ for the SB0 galaxy NGC~936.  Indirect 
estimates can be obtained by matching hydrodynamical simulations with 
observations; Athanassoula's (1992) comparison of shocks in 2-D gas flows with 
the locations and shapes of dust lanes in barred galaxies led her to suggest 
that $\lag/\len = 1.2 \pm 0.2$.  Studies of specific galaxies (Lindblad et al.\ 
1996; Weiner et al.\ 1996), using potentials derived from the observed near 
infrared light distribution, yield similar values.  While the evidence is 
meager, it does suggest that $\lag/\len \ltsim 1.5$ for Hubble types SB0 through 
SBc.

Our simulations show that, as predicted, bars rotating inside a halo of moderate 
central density slow down in a short time, yielding $\lag/\len > 1.5$.  It seems 
possible to avoid an inconsistency with the observed properties of galaxies only 
by reducing the halo central density to the point where the luminous matter 
dominates the rotation curve throughout most of the disk.

\section{Numerical Technique}
We create fully self-consistent $N$-body models of disk-halo galaxies to 
simulate their evolution.  We have employed two different types of grid, having 
both Cartesian and cylindrical polar geometries, and upwards of $300\,000$ 
particles.  We have verified that results are insensitive to the numerical 
scheme and are unchanged when $N$ is increased by a factor of five.

\ifodd\style
  \begin{figure}[t]
  \centerline{\psfig{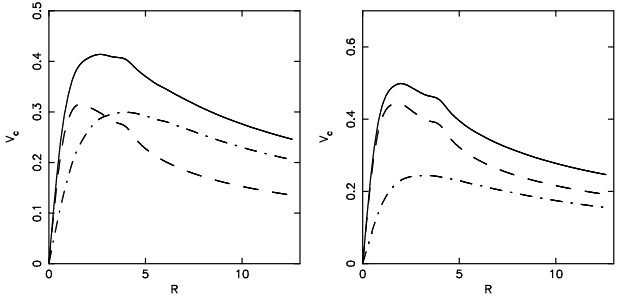}}
  \caption{Rotation curves for the most massive halo (left) and the least
  massive halo (right) from the series of models without extensive halos.
  The curves are decomposed into disk (dashed) and halo (dot-dashed)
  contributions.  The total mass is the same in both models; the disk mass
  fractions are respectively 0.3 and 0.6 in the left and right panels.}
  \end{figure}
\fi

Our initially axisymmetric equilibrium models were created by an extension of 
the procedure first employed by Raha et al.\ (1991) that takes into account the 
combined gravitational fields of the disk and halo.  The halo particles in all 
the models we present in this {\it Letter\/} are selected from a non-rotating, 
isotropic distribution function (DF) of polytropic form.  The initial disks have 
velocity dispersions set by adopting a constant Toomre $Q$ parameter and 
thickness at all radii.  A fuller description of our numerical procedures will 
be given in a later paper.

\section{Simulation Results}
To illustrate our main point, we first present a series of models with halos 
that do not extend to large radii.  These experiments were run on modest grids, 
which cannot contain extensive halos if they are to afford reasonable resolution 
for the disk.  We later present a model run on a much larger grid to demonstrate 
that a more realistic halo does not alter our conclusion.

\subsection{Models without extensive halos}
The rotation curves for two models, together with the separate contributions of 
the disk and halo, are shown in Figure 1.  These models have the greatest and 
least massive ``halos'' of a series of four models run on moderate size grids; 
the only property that varies along this sequence is the disk/halo mass ratio.  
In none of the models is the halo dense enough to suppress the bar instability 
in the disk.  All quickly form strong, rapidly rotating bars which subsequently 
bend and thicken in the usual way (Combes et al.\ 1990, Raha et al.\ 1991).

As predicted, as soon as the disk is strongly non-axisymmetric, the bars begin 
to lose angular momentum through dynamical friction.  The rotating bar quickly 
induces a bisymmetric distortion in the distribution of halo particles which 
lags the bar by up to $\sim 45^\circ$; the consequent torque gradually decreases 
as the distortion in the halo becomes aligned with the bar.

In the model with the most massive halo, angular momentum transfer ceases after 
about 50 initial rotation periods of the bar, by which time the halo has gained 
some 40\% of the disk's initial angular momentum.  As the bar loses angular 
momentum, its pattern speed drops -- by a factor of five in this model -- in 
agreement with Weinberg's (1985) conclusion that bars can be rapidly braked by 
dynamical friction.

The bar in the model with the least massive halo, on the other hand, continues 
to rotate quite rapidly.  Bar deceleration is less for two reasons: (a) a rather 
mild distortion in the lower density halo has a major axis which lags the bar by 
only $\sim 10^\circ$ creating a much weaker torque, and (b) the disk, and 
therefore the bar, mass is greater so that the pattern speed changes produced by 
the weaker torque are much less. 

\ifodd\style
  \begin{figure}[t]
  \centerline{\psfig{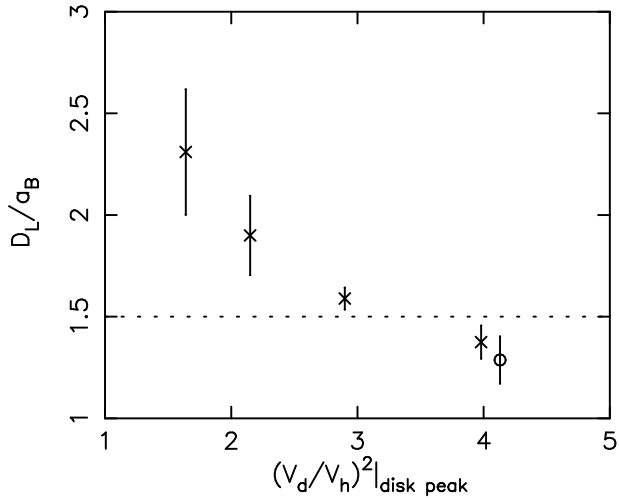}}
  \caption{The final equilibrium $\lag/\len$ plotted against the square of the
  initial ratio of disk to halo circular velocities, evaluated at the peak of
  the disk rotation curve.  The error bars are calculated from time averages.
  The dotted horizontal line is at $\lag/\len = 1.5$.  The different symbols
  are explained in the text.}
  \end{figure}
\fi

The ratio $\lag/\len$ increases steadily as the bars slow down.  The crosses in 
Figure 2 show our main result from this series of models, that the final value 
of $\lag/\len$ increases systematically with the mass of the halo.  Only our 
model with the extreme low mass halo has $\lag/\len \ltsim 1.5$, and remains in 
the range consistent with the observed values discussed in the Introduction.

\subsection{A more realistic model}
Since these models have unrealistic halos, we have run a further experiment on a 
much larger grid in which the rotation curve is flat over a wide range of radii, 
as shown in Figure 3.  Once again the halo was a simple, non-rotating polytrope 
with an isotropic distribution of velocities, but here we adjusted the halo mass 
and outer cut-off radius to give us a maximum disk model.

The behavior of this much more realistic case is very similar to that of the low 
mass halo model from the earlier sequence.  The disk quickly formed a strong bar 
which slowed only slightly; the bar pattern speed stabilized some 25\% below its 
initial value as angular momentum loss to the halo dropped to a very low rate.  
The ratio $\lag/\len$, shown by the circle in Figure 2, settled to about 1.3 
after just three bar periods and remained at this value for the next 12 bar 
rotations with no hint of an increase.

\ifodd\style
  \begin{figure}[t]
  \centerline{\psfig{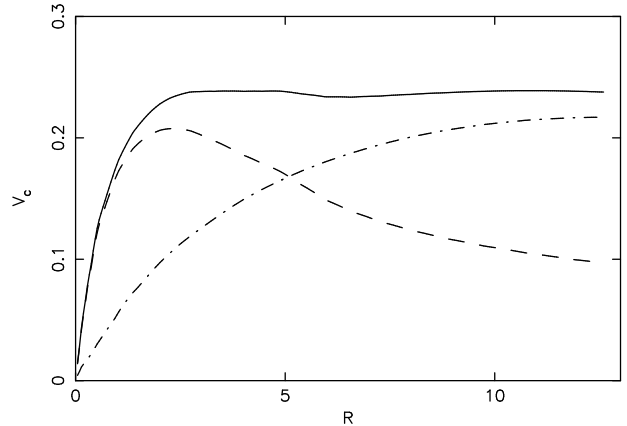}}
  \caption{The rotation curve for the maximum disk model decomposed into disk
  (dashed) and halo (dot-dashed) contributions.}
  \end{figure}
\fi

The central halo density in this model is low; at the peak of the disk rotation 
curve $V_{\rm disk}/V_{\rm halo}$ is about the same as in the model with the 
lowest density halo in the sequence described in \S3.1.  Since the resulting 
value for $\lag/\len$ is also quite similar, we argue that 
dynamical friction is determined by the density of the
inner halo and not the total halo mass.  This conclusion is in
agreement with previous wisdom (Binney \& Tremaine 1987, \S7.1) that
dynamical friction in an inhomogeneous stellar system depends upon the
local mass density at the position of the perturber.

It should be noted that our chosen parameter, $(V_{\rm disk}/V_{\rm halo})^2$
at some radius is a measure of the interior halo mass and not the local 
density.  The critical value may therefore depend upon the density profile of
the inner halo.

\subsection{Other models}
We have run many more models to experiment with rotating and anisotropic halos, 
and have found that friction is not lessened dramatically for any reasonable 
halo DF.  Making more halo particles orbit in the same sense as the disk {\it 
does\/} reduce friction, but even giving almost all halo particles the same sign 
of angular momentum still led eventually to an unacceptably slow bar in our 
densest halo.  
In no case were the effects of secondary bar growth (Sellwood 1981) enough to 
compensate for the bar deceleration.  We give further details of these tests 
elsewhere.

\section{Discussion}
The stellar dynamical models just described exclude possible mechanisms that 
could speed up bars.  Gas is driven inwards by angular momentum loss to the bar, 
but it seems unrealistic to suggest that enough could be gained from gas to 
compensate for the 40\% lost by the disk in our densest halo case.  More 
importantly, mass inflows lead to a secular increase in the central density 
which will cause the pattern speed to rise (Kalnajs 1996).  Taken to excess, 
however, too great a mass build up in the center will destroy a bar (Norman et 
al.\ 1996).  Note also that gas driven changes of this kind cannot occur in SB0 
galaxies such as NGC 936 which is known to have a fast bar.

If bars were short-lived, they could dissolve before they have time to slow 
down.  We do not regard this idea as attractive for a number of reasons.  First, 
to avoid any slow cases, bars would have to be destroyed relatively soon  
after they formed.  Second, unless all bars have recently formed for the first 
time (implausible), the large fraction of disk galaxies that are strongly barred 
($\sim 30\%$, Sellwood \& Wilkinson 1993) would require short-lived bars to 
recur in every galaxy.  Bar dissolution (through a growing central mass or a 
minor merger, say) would leave the stellar disk hot and unresponsive and it seems 
inconceivable that disks could repeatedly recover to allow rapid redevelopment 
of fresh bars.  Third, the bars in all our simulations are very robust and 
survive even a five-fold reduction in pattern speed.

Weinberg (1985) suggested that the disk may be able to transfer angular momentum 
to the bar at an inner Lindblad resonance, keeping it in rapid rotation.  
Previous simulations (e.g.\ Sellwood 1981), as well as those described here, 
show that the net exchange of angular momentum between the bar and the disk 
amounts to a loss by the bar.  It does not seem possible that a rapidly rotating 
bulge could torque up the bar either: even when we gave near maximal rotation to 
our densest halo simulation, in which some fraction of the halo mass could be 
thought of as a rapidly rotating bulge, the bar still slowed unacceptably.

\subsection{The Milky Way}
The contribution of DM to the circular velocity at the Solar position is still a 
matter of debate (Kuijken \& Gilmore 1991; Merrifield 1993; Sellwood \& Sanders 
1988; Sackett 1997).  There is also some disagreement over the precise size 
(Weinberg 1992; Dwek et al.\ 1995) and pattern speed of the bar (Binney et al.\ 
1991; Kalnajs 1996), but it seems unlikely that the value of $\lag/\len$ will 
turn out to be substantially larger than for other barred galaxies.  
Verification that the bar in the Milky Way is indeed fast would strengthen the 
arguments in favor of a maximum disk.

\subsection{Unbarred galaxies}
We have argued that barred galaxies must have maximum disks in order that the 
bar rotation rate remain consistent with the believed properties of barred 
galaxies.  Here we extend the argument to unbarred galaxies.  There are no 
qualitative differences between the outer rotation curve shapes of SB and of SA 
galaxies (Bosma 1996), which suggests that their DM mass distributions are quite 
similar.  Ostriker \& Peebles (1973) argued that the survival of an unbarred 
disk requires a galaxy to be DM dominated, implying that the disks in SA 
galaxies should be less than maximum.  If we accept their argument for the 
moment, and also the seemingly reasonable proposition that there is a continuum 
of DM content from SA to SB galaxies, then we are forced to conclude that there 
should be some barred galaxies with enough halo to brake the bar but not enough 
to suppress it, as in our models.  The apparent absence of slow bars in high 
surface brightness galaxies implies that there is no continuum between high and 
low density halos.  We find a single, maximum-disk distribution to be more 
natural than a bimodal distribution of halo fractions, and some other mechanism 
must be responsible for the stability of unbarred galaxies (e.g., Binney \& 
Tremaine 1987, \S6.3.2).

\subsection{Prospects for finding slow bars}
The newly discovered class of low surface brightness (LSB) galaxies (see Bothun, 
Impey \& McGaugh 1997 for a review) exhibits the same Tully-Fisher relation as 
do high surface brightness (HSB) galaxies (Zwaan et al.\ 1995), and therefore 
must contain larger fractions of DM (de Blok \& McGaugh 1996).  Any bars in 
galaxies of low or intermediate surface brightness with significant DM content 
should have been slowed by dynamical friction.  The absence of known slow bars 
may simply reflect an observational bias in favor of HSB galaxies.

\section{Conclusion}
We have argued that the apparent absence of slow bars in HSB galaxies indicates 
that DM should not have a high central density.  We find that dynamical friction 
from slowly rotating, isoptropic halos is unacceptably strong when the halo
makes a major contribution to the circular speed at about two disk scale 
lengths -- i.e., within the region where bars are 
found.  It is clearly desirable to determine pattern speeds of many more bars, 
and we suggest that promising candidates for slow bars might be found amongst 
galaxies of low or intermediate surface brightness.

\bigskip
We thank an anonymous referee for constructive comments.  This work was 
supported by NSF grant AST 93/18617 and NASA Theory grant NAG 5-2803.

\ifodd\style
\else
\newpage

\figcaption{Rotation curves for the most massive halo (left) and the least 
massive halo (right) from the series of models without extensive halos.  The 
curves are decomposed into disk (dashed) and halo (dot-dashed) contributions.  
The total mass is the same in both models; the disk mass fractions are 
respectively 0.3 and 0.6 in the left and right panels.}

\figcaption{The final equilibrium $\lag/\len$ plotted against the square of the 
initial ratio of disk to halo circular velocities, evaluated at the peak of the 
disk rotation curve.  The error bars are calculated from time averages.  The 
dotted horizontal line is at $\lag/\len = 1.5$.  The different symbols are 
explained in the text.}

\figcaption{The rotation curve for the maximum disk model decomposed into disk 
(dashed) and halo (dot-dashed) contributions.}

\fi

\end{document}